# Machina Economicus: A New Paradigm for Prosumers in the Energy Internet of Smart Cities

**Luyang Hou[1*], Jun Yan[2], Yuankai Wu[3], Chun Wang[2], Tie Qiu[4]**

[1] *School of Computer Science (National Pilot Software Engineering School), Beijing University of Posts and Telecommunications, 10 Xitucheng Rd, Haidian District, Beijing, 100876, China.*
*Corresponding author E-mail: Luyang.hou@bupt.edu.cn*

[2] *Institute for Information Systems Engineering, Concordia University, 1455 De Maisonneuve Blvd. W., Montréal, H3G 1M8, Québec, Canada*

[3] *College of Computer Science, Sichuan University, No. 29, Wangjiang Road, Wuhou District, Chengdu, Sichuan, 61006, China*

[4] *School of Computer Science and Technology, College of Intelligence and Computing, Tianjin University, Tianjin, 300350, China*

**Abstract:**
Energy Internet (EI) is emerging as new share economy platform for flexible local energy supplies in smart cities. Empowered by the Internet-of-Things (IoT) and Artificial Intelligence (AI), EI aims to unlock peer-to-peer energy trading and sharing among prosumers, who can adeptly switch roles between providers and consumers in localized energy markets with rooftop photovoltaic panels, vehicle-to-everything technologies, packetized energy management, etc. The integration of prosumers in EI, however, will encounter many challenges in modelling, analyzing, and designing an efficient, economic, and social-optimal platform for energy sharing, calling for advanced AI/IoT-based solutions to resource optimization, information exchange, and interaction protocols in the context of the share economy. In this study, we aim to introduce a recently emerged paradigm, Machina Economicus, to investigate the economic rationality in modelling, analysis, and optimization of AI/IoT-based EI prosumer behaviors. The new paradigm, built upon the theory of machine learning and mechanism design, will offer new angles to investigate the selfishness of AI through a game-theoretic perspective, revealing potential competition and collaborations resulting from the self-adaptive learning and decision-making capacity. This study will focus on how the introduction of AI will reshape prosumer behaviors on the EI, and how this paradigm will reveal new research questions and directions when AI meets the share economy. With an extensive case analysis in the literature, we will also shed light on potential solutions for advancements of AI in future smart cities.

**Abbreviations**

| | |
|---|---|
| **AI** | **Artificial Intelligence** |
| **AV** | **Autonomous Vehicle** |
| **BEV** | **Battery Electric Vehicle** |
| **D³** | **Decentralized, Dynamic, Data-driven** |
| **DER** | **Distributed Energy Resource** |
| **DR** | **Demand Response** |
| **EI** | **Energy Internet** |
| **EV** | **Electric Vehicle** |
| **FCEV** | **Fuel Cell Electric Vehicle** |
| **GHG** | **Greenhouse Gas** |
| **ICT** | **Information and Communications Technology** |
| **IoT** | **Internet of Things** |
| **ITS** | **Intelligent Transportation System** |
| **MARL** | **Multiagent Reinforcement Learning** |
| **MAS** | **Multiagent System** |
| **RES** | **Renewable Energy Sources** |
| **RL** | **Reinforcement Learning** |
| **SG** | **Smart Grid** |
| **V2G** | **Vehicle-to-Grid** |

1. Introduction

Global climate challenges have become increasingly impactful, and developing a low-carbon economy has become a global trend in smart cities. Modern smart cities have been seeking sustainable and efficient innovations in energy management, transportation, health care, governance, and other sectors to meet the sustainable development goals of urbanization (Silva et al., 2018). The transformation towards 'carbon neutrality' and 'net-zero emission' requires minimizing the use of fossil carbon in energy production, transportation movements, manufacture and, not least, energy utilities and transmission networks themselves (Selman, 2010). Meanwhile, decarbonization requires shifting to low- or zero-emission energy carriers at the final point of consumption (i.e., electricity, hydrogen, or bioenergy) and a clean energy supply (O'Malley et al., 2020). Energy Internet (EI), perceived as an integrated information and energy architecture, combines 'green' features with emerging technologies such as cloud computing, Internet-of-Things (IoT), big data analytics, and edge intelligence, among others, to utilize diverse distributed energy resources (DERs) and the abundant energy big data (Wang et al., 2018; Zhou et al., 2016). EI promotes the formation of more flexible, personalized, and efficient energy production and consumption such that the energy flow can be incorporated with information, value, and business flows. The advancement of EI, with renewable energy, power electronics, and information and communication technologies (ICT), also drives the transition of traditional electricity consumers to energy prosumers.

The operations and management of EI involve various technological drivers and broad research issues including, but not limited to, big data analysis, resource allocation, integrated energy services, microgrid management, demand side management, and information exchange, among others. As a complex system, EI also incorporates uncertainties, nonlinearities, the need for pervasive sensors, and coordinated, cooperative, distributed control and optimization. To this end, artificial intelligence (AI) is invaluable for its role in this transformational stage facing EI as the field that studies the synthesis and analysis of computational agents that act intelligently (Poole & Mackworth, 2010). AI is already at the forefront of driving valuable strategies to optimize our society across all operations. The advent and impacts of new technologies and big data accompanied by AI are evidencing the benefits of its implementation in facilitating missions for achieving smart and sustainable-infrastructure initiatives and net-zero commitments. Thereby, EI allows AI to accelerate the development of smart grid (SG) and intelligent transportation systems (ITS) in smart cities driven by sustainability goals and motivations or actions of heterogeneous economic stakeholders, such as DERs, electric vehicles (EVs), electrolyzers, charging/refuelling stations, data centers, among many others.

In general, EI can be modelled by multiagent systems (MASs) derived from distributed artificial intelligence (M. Wooldridge, 2009). In a MAS, the stakeholders of EI are deemed as homo economicus agents, who are perfectly rational, and their private information is single-dimensional, statistically known or does not change over time (Tang, 2017). However, none of these assumptions hold in the practices of EI, where agents may have different levels of rationality, are not directly coordinated or controlled by a central authority and have private information that may also change over time. More importantly, perfect rationality is unachievable with finite computational resources, and the prediction of the effects of machine learning algorithms on humanity poses a substantial challenge (Rahwan et al., 2019). Paul R. Milgrom and Robert B. Wilson, laureates of the 2020 Nobel Economic Sciences Prizes (Dynamics & of AI Systems in Complex Networks, n.d.), also pointed out that an economic analysis on homo economicus is difficult, as they behave strategically based on the available information, only considering what they know themselves and what they believe others to know.

In this context, it is crucial to understand various dimensions for EI modelling, particularly where systems may be optimized for specific contexts with three distinguishing aspects standing out. First, the multi-modal connectivity and system interoperability in EI requires the integration of agents' feedback into the decision-making loop. Fields of microeconomic theories, especially game theory and mechanism design, provide useful tools and examples of the design of efficient market mechanisms (Hou & Wang, 2017). Second, the massive data exchange between EI subsystems requires extensive communication efforts, thus making systems execute at high frequencies where the parameters should be flexibly adjusted to adapt to the dynamics. On the other, the uncertain, non-statistically known payoffs and the coordination of agents remain to be solved in a decentralized, dynamic, and data-driven ($D^3$) environment. Moreover, the probability density function of data may also change over time in nonstationary environments, particularly those with multi-domain data (Ditzler et al., 2015). Third, designing a solution framework needs to address the stochastic process governing the agent's feedback and preferences over time. Restricted by computing ability and accessible data, the classic economic theory only applies in very restricted settings under strong assumptions of the agent's full rationality and formulaic payoff functions. The equilibrium-based solutions are thereby not always optimal from the system-level perspective. In real-world situations, the agents may have different level of rationality and risk sensation; their preferences may also change over time, rendering related assumptions in common mechanisms ineffective (Sandholm, 2003). In summary, a core challenge to EI is the trade-off between two objectives: system-wide efficiency and - at the same time - individual well-being with the agents' selfishness and collective behaviors. Economic efficiency can only be achieved only if the outcome maximizes the sum of the utilities of the agents (Poole & Mackworth, 2010).

To this end, the integration of different components is gradually associated with information gathering and decision-making by AI - instead of human agents - through a coordinated, cooperative, and distributed manner. With the increasing adoption of AIs, machine-based interactions have been extensively explored with learning-based approaches (Vandael et al., 2015, Sedighizadeh et al., 2019). However, the agents' self-oriented and myopic views may degrade the system-wide efficiency, where the synchronous strategy learned by these independent AIs may cause bad equilibria in society. Understanding AI's social and economic impact is necessary, especially when energy systems are becoming more integrated and controlled by AIs or aggregating information held by AIs, as experienced in autonomous driving or energy market trading. In the near future, selfish, utilitarian and human-conscious AI ecosystems will begin to emerge in our society, where the AIs can estimate the payoff gain or loss in their adoption (Fernandes et al., 2020). Following this trend, AIs will gradually replace human decision-making, respond rationally to others' behaviors, and interact with each other as in homo economicus, which calls for establishing interaction rules for these artificial and economic agents to create synergies between machine learning and economic theory. As a result, AIs should not just approach homo economicus-like ability, but evolve into an entirely new species of economic agent: ***Machina Economicus***, which is proposed by D. Parkes and M. Wellman (Parkes & Wellman, 2015).

The key contribution of this study is to present a cutting-edge economic-AI perspective to the management and operation issues of the EI. Such economic-AI perspective is manifold and significant: a market perspective can shape AIs behaviors and unlock economic opportunities for decision-makers; furthermore, AI techniques will facilitate the evolution from manual system design to automated and data-driven design when gathering, distributing, storing, and mining data and state information in the EI. Given this, we attempt to fill the gap between the paradigm of machina economicus and key research issues in EI by providing detailed explications of

- a review of EI relevant research issues in the existing literature, and the research gaps in EI, as well as the energy prosumers concept in EI;
- an investigation on machina economicus-centered EI architecture, a thorough analysis of the characteristics of machina economicus and its applicability to EI;
- a review of the state-of-the-art models and methodologies that integrate mechanism design, machine learning, and optimization, and an integrated solution and a step-by-step methodology that extends these approaches to machina economicus-centered EI architecture.

## 2. Literature Review

Energy Internet covers numerous topics in the smart city that include sustainable energy systems, data collection and management, infrastructures and relevant technologies, urban transportation mobility, information transmission and communications, and collaborative economy, among many others. This section only focuses on some key applications and research issues related to EI. We will examine how machina economicus interact with both individuals and entities to boost the capabilities of EI and influence social, economic, and systematic outcomes. Notably, the operations and management of EI is a multidisciplinary field of operations research, economics, and computer science, which requires knowledge from various smart structures, interconnected networks, and big data to elicit economic issues in system operations. Designing a machina economicus-centered EI needs understanding and analyzing the structural properties and critical features in EI. In what follows, we will introduce several research applications and exemplify some key challenges from adopting machina economicus in EI.

**2.1 Key Applications**
**2.1.1 Energy management in smart grid**

A smart grid (SG) can be seen as an electric system that applies information, two-way communication technologies, and computational intelligence in an integrated way across electricity generation, transmission, substations, distribution, and consumption (Fang et al., 2011). The smart grid provides the capacity to manage vastly distributed generation, enabling large quantities of renewable sources and local microgrids (Masera et al., 2018). Energy demands tend to peak at some specific times of the day depending on the needs of industrial, commercial, and domestic consumers, which may bring higher costs to end-users and instabilities to the electricity networks (Bandyopadhyay et al., 2016). Microgrids are advancing the management efficiency and security of power grids by integrating DERs, energy storage systems and distributed controllers (Yolda\cs et al., 2017). Electric vehicles (EVs), driven by the emission reduction goals, will contribute significantly to future load peaks in microgrids, affected heavily by the uncertain arrivals of EVs and their charging demands. However, the objectives of EVs and SG actors are not always aligned, where users want to take quick and economical charge, and power providers are driven by profits and grid stability. Managing EVs' unidirectional energy flows (or bidirectional in future scenarios) requires new economic models that can effectively incentivize user participation, following the grid's control and operation needs (Shuai et al., 2016). Another important factor is energy storage systems with large, distributed renewable energies. Storage devices can capture energy during periods of high renewable energy production and release stored energy when needed, reducing both the quantity of conventional generation resources required to mitigate variability in renewables production and the total amount of energy curtailed (Silva-Monroy & Watson, 2014).

**2.1.2 Demand response (DR)**

Demand response is considered the most cost-effective and reliable solution for smoothing the demand curve in smart grids (Vardakas et al., 2015). The key idea is to motivate the changes in the power consumption habits of consumers in response to incentives regarding electricity prices. DR programs mostly apply to residential, commercial, and industrial users through time-varying prices or incentives. The objective is to reduce the total power consumption and generation and eliminate overloads in distribution systems, especially in the context of zero-emission buildings (Umetani et al., 2017) or smart homes (Thapa et al., 2017). Moreover, the connection of thousands of EVs to the grid at one time will be essentially a massive backup battery for power generators, and the vehicle-to-grid (V2G) paradigm can better use RES. Hou et al., (2020) presented a reinforcement learning-based pricing mechanism to determine the optimal charging prices considering users' price sensitivity and individual preferences, where the strategic interaction between the charging station and electricity users is modelled as a discrete finite Markov decision process.

### 2.1.3 V2X (Vehicle-to-X)

The high penetration of EVs motivates the power grid to provide more incentives for users to adjust the timing of charging. V2X paradigm, i.e., vehicle-to-grid (V2G), vehicle-to-vehicle (V2V), vehicle-to-building (V2B), or vehicle-to-home (V2H), unlocks further flexibility potentials in microgrids (Outlook, 2020). These technologies enable a smoother integration of EVs with power systems and variable RES generations, allowing EVs and DERs to actively participate in the energy exchange process, with their privacy and preferences being respected. To achieve systematic operational efficiency, a framework of ahead-of-time scheduling and real-time control can integrate photovoltaic charging station supply, charging and discharging of EVs, and real-time energy management (Yan et al., 2019). Furthermore, the incentive-based mechanisms should be designed to make electricity prosumers understand the operational implications of, and agree to, autonomously chosen trading decisions and participate in energy markets. On top of the existing literature, a multi-layer sequential decision-making framework for energy management can jointly solve electricity purchasing and dynamic pricing of charging stations and charge/discharge control of EVs. The multi-stage strategic interaction of station-station, station-user, and user-user is expected to achieve common good solutions with efficient interacting rules and machine learning algorithms.

### 2.1.4 Integrated hydrogen-electrical energy system

Hydrogen, as a low-carbon fuel, is receiving more and more attention because it can link renewable resources to scalable services. As a low-carbon fuel, hydrogen is light, storable, energy-dense, and produces no direct emissions of pollutants or GHG (IEA, 2019), which can greatly reduce GHG emissions. In particular, green or renewable hydrogen – made from the electrolysis of water powered by RES, such as solar or wind – is indispensable to climate neutrality (van

Renssen, 2020). An integrated energy system architecture exploring critical links among energy, transportation, ICT facilities, and urban design was proposed by Schröder et al., (2020). The architecture owns zero-emission vehicles, integrated RES, energy storage and dispatching systems. The photovoltaic array generates an electric current that charges EVs and is directed to water electrolysis to produce hydrogen; the intermittent solar energy can also be stored in the intelligent bi-directional charging system. In addition, hydrogen can also be injected into the natural gas grid to lower GHG emissions. Fuel cell electric vehicles (FCEVs) are connected to the Dutch national grid (Robledo et al., 2018), allowing fleets to deliver up to 10 kW direct current power output. Such an FCEV2G system could reduce electricity imported from the grid and obtain a more self-sufficient all-electric energy loop. The researchers have already begun to investigate the co-management of hydrogen electrolysis with other distributed renewable energy sources in a local distribution network or a microgrid (Hou et al., 2022). Despite the advantages of hydrogen, however, a real-world concern for a large-scale deployment of hydrogen system in the EI is the requirements of significant capital investment and long-term commitment, but faces high risks of low demand, and poor short-term returns.

### 2.1.5 Blockchain-based peer-to-peer energy trading

The electricity market relying on centralized generation from utility providers has started to transform into a decentralized and dynamic market, facilitating a new paradigm of peer-to-peer (P2P) energy trading among DERs. This new paradigm raises many new challenges, among which this proposed project will mainly focus on two issues: the trustworthiness of distributed energy transactions and the precise control of distributed energy management, which are critical factors for social well-being in EI. Data collection from users or IoT devices will raise many ethical issues and concerns, such as how to manage numerous privacy implications, how heterogeneous data sources work together, etc. The information security associated with the roles of humans in the security process and the cyber security with confidentiality, integrity and availability issues should be carefully addressed (Von Solms & Van Niekerk, 2013). These dimensions have ethical implications for society. For instance, cyber security is a critical issue in SGs, as millions of electronic devices are interconnected via communication networks throughout critical power facilities (Fang et al., 2011). The connectivity of smart energy systems enables personal information collection, which may invade users' privacy. Blockchain is a sort of distributed ledger technology and has features of decentralization, anti-tamper, transparency, anonymity, and contract autonomy (Huo et al., 2022). These features help improve the services and promote the development of EI. Energy prosumers would like to participate in energy transactions: for example, a decentralized approach based on transactive energy systems and P2P energy transactions could be adopted to effectively manage virtual power plants (Siano et al., 2019).

**2.2 Research gaps with energy prosumers**

Decarbonizing transport and energy systems require switching to clean alternative fuels (e.g., sustainable biofuel, electricity, or hydrogen produced from renewable sources) and/or increasing vehicle efficiency (O'Malley et al., 2020). Smart energy and electricity networks are crucial components in building a smart city, as they not only can facilitate the integration of RES and the electrification of transportation but also enable new energy-related value-added services (Masera et al., 2018). However, it is challenging to match DERs with diverse energy demands in practical scenarios, and distributed battery storage can hardly achieve large-scale inter-seasonal power regulation due to its limited capacity and concerns about battery degradation. A potential solution is to integrate transport, power, and hydrogen systems and build efficient coordination strategy to balance demands with supplies. Instead of passively consuming power by observing tariffs from utilities, EI can support energy prosumers, who can actively trade energy and ancillary services as well as set local energy prices through a connected platform (Parag & Sovacool, 2016). Prosumers can also participate in dynamic pricing, acting both as the price taker and maker. Compared with sharing rides or vehicles, sharing building-based energy management creates richer content to explore (Qi & Shen, 2019). Such sharing-enabling entities can make decisions on electricity pricing and matching to incentivize prosumer participation and coordination, as well as ensure grid stability. Moreover, it will transform passive assets, such as electric or fuel cell vehicles, charging stations, and renewable energies, into active participants in microgrid energy transactions.

Most DR programs encourage slow and off-peak charging, which requires a continuing connection with power systems. However, DR programs that simply rely on reacting to control or price signals will not be enough to deal with decision-makers' selfishness in $D^3$ environments, which makes it hard to apply to taxi fleets, logistics vehicles, or public transport. Consumers' selfish nature and their dynamic demands and preferences should not be neglected in seeking economic benefits and implementing social objectives rather than just reducing their charging costs. Another challenge comes from decision interdependency. Since all stations solely perform dynamic pricing based on local observations and update their decisions simultaneously, the pricing strategy for a single station depends on the settlement of other stations. Therefore, building dedicated charging facilities and designing effective negotiation protocols can introduce more flexibility to energy management.

As transportation is responsible for a large portion of GHG emissions – especially for heavy-duty vehicles – emissions and other environmental factors should be integrated with transportation efficiency objectives. Transportation electrification provides more opportunities for low/zero-carbon vehicles, which should be incorporated with government policy goals, land use planning, urban design, and associated system management as integral components of the overall system design and modelling strategies. To solve this problem effectively, improve transportation efficiency, and reduce energy consumption, effective methods, such as multi-objective optimization, are required to examine these two contradictory objectives. As the integration of shared vehicles scales up in city-scale transportation, it is necessary to understand the management strategy and implications of such hybrid business model. The following questions remain to be explored in the current EI environments: *what are the impacts of integrating such electric mobility with EI on the pricing/dispatching decisions of multi-side platforms*? *How to integrate AI agents' economic actions into the interaction of multiple stakeholders in such distributed and data-driven environments*?

Key to answering these questions is to characterize the responsiveness of prosumers, the considerations of system dynamics, and the effective negotiation or communication protocols. In addition, big data needs to be accommodated by transforming decision-making to be more data-driven. Most leading methodologies for management and operational issues in EI have focused on optimizing operational costs or maximizing RES usage, limiting our ability to answer questions about how energy prosumers and society-at-large share various and limited distributive resources in a fair way. The social welfare perspective should directly engage system efficiency, fairness, and individual payoffs in the design of optimization strategies. Social welfare is usually defined as a weighted sum of individual utilities (Hu & Chen, 2020). Manipulating weights will significantly alter the outcomes, which is also the way to compute optimal (fairness-constrained) allocations under a set of welfare weights. How to define fairness as a metric or a property of the optimization objective or constraints at the intersection of computing and economics is worth exploring in EI. An interactive auction market with a bidding system could be solvable, where a coordinator can accept or reject a prosumer's bids to match demand and supply, manage demand at peak times, and improve social welfare.

**3. Machina Economics Paradigm**

The synergetic development of machina economicus and EI is raising numerous promising practices in $D^3$ environments. The design for machina economicus ecosystems will admit more complex interfaces in ITS and SGs in smart cities, which will simultaneously provide opportunities with machine learning, innovative collective intelligence, interaction rules in game theory, and research thrusts to machina economicus-centered infrastructure. One realm where AI's inroads are conspicuously haphazard is economics, much of this stemming from embedded ideological commitments. These predilections make it exceptionally difficult for the mainstream discipline to fully embrace the greater implications of AI found in the study of the economy as a complex adaptive system (Daneke, 2020). Classic economics assumes that homo economicus always make the best and most rational decisions based on perfect knowledge about their economic actions. Under this assumption, game theory studies and predicts how self-interested agents behave in the MAS and derive general equilibrium solutions. Reversely, mechanism design works with the presence of self-interested homo economicus that the socially desirable outcomes arise naturally from the strategic interaction. However, researchers have gradually begun to realize that humans are not rational enough in real-world practices and lack the necessary abilities to make informed decisions about how to obtain the maximum of their own benefits. An equilibrium solution is not the end game to satisfying a welfare goal, as it often assumes preferences from the population and does not directly relate to the actual outcomes that such a population will receive.

AI can mimic homo economicus if it can align perceptions, preferences, and actions to come to decision-making under uncertainty (Martin, n.d.). Now humans have overcome their deficiencies by using tools and have developed machines to allow us to inch closer to the idealized economically perfect agents. Their decision-making ability can be greatly amplified with the assistance of machine intelligence when environments are generating a large amount of information and data that are available to base their actions on. Obtaining an insight into the behavior of such AI systems is essential to control their actions, reap their benefits, and integrate them with smart infrastructures (Rahwan et al., 2019).

Machina economicus is an AI agent with a coupling of perception, reasoning, and acting, which lies in the intersection of AI and economics. They can better respect the idealized assumptions of rationality made and help improve the fairness in the economic systems than homo economicus, and they own much stronger abilities of computation and decision-making. These selfish AIs adopt a game-theoretic view of the world, where agents rationally respond to each others' behavior, presumed (recursively) to be rational as well. In addition to the autonomy, reactivity, social ability, and pro-activeness properties shared by the intelligent agents (M. J. Wooldridge & Jennings, 1995), machina economicus displays an economic view and human-like attributes with their learning ability that can best approximate the economic rationality. Specifically, (1) they are capable of perceiving the world around them by replacing human decision makers with machines, taking actions and making decisions to advance specified goals; (2) they can process more information and make more reasonable decisions and, at the same time, anticipate negative consequences and interact more effectively with others; (3) they have the ability to learn how to aggregate individual preferences and economic rationality into collective decision-making, which enables them to deepen the understanding by seeing through the perspectives of others and it highlights what may constantly be overlooked when refining and enhancing decision-making.

A machina economicus-EI architecture can be perceived as a two-layer interconnected network that provides data, and information-energy flow about the state of the systems for decision-making, optimization, and management. Such an architecture consists of a strategic and physical layer, where the physical layer consists of a set of components, modules, or ICTs. The strategic layer studies how these components interact and make decisions by collecting and integrating data from microgrids, public transport, buildings, parking places, zero-emission vehicles, various sensors, and/or ICT facilities. Under the IoT architecture, smart networks collect and monitor data through the gateways on a large scale, and smart devices utilize appropriate communication methods and cloud services to integrate digital information and physical objects, which provide machina economicus with data and services for decision-making (Qiu et al., 2018). Over the physical layer, the strategical layer acts more like the brain taking charge of data analysis and the decision-making about competition, collaboration, or cooperation with others, which can be implemented in a physical device. Machina economicus has its goals or preferences over states of the environment. They can analyze the data and adopt the corresponding strategy to improve its behaviors and efficiency in an offline or online manner, rather than just breaking down the decisions into primitive operations. In EI, machina economicus can cover electric power grids, IoT devices and smart facilities, electric/autonomous vehicles, DERs, RESs, and organizations, which may all operate under the new paradigm.

Besides, machina economicus might be designed, deployed, owned, and/or operated by a myriad of parties. Market operations should adjust to a larger variety of heterogeneous entities and allocate resources in a fair and Nash-optimal manner (Conitzer et al., 2019). These selfish AIs have different types (Agarwal et al., 2020);

therefore, addressing their conflicting goals requires not only effective learning algorithms but also proficient interaction rules and negotiation protocols. A simultaneous auction framework allows agents to interact in different markets and compete for multiple resources. In addition, there is a potential to transfer models, data, and knowledge across heterogeneous markets in terms of the insufficiency of data and domain of interest, which will greatly improve the performance of learning by avoiding many expensive data-labeling efforts (Pan & Yang, 2009). The traditional way to model agents' preferences is to use a generalized logarithmic payoff model as the premier model of financial markets in economic theory (Rubinstein, 1977). The key challenge is that the distribution of agents' preferences always changes and varies in different real-world applications, which makes the logarithmic model inaccurate and the collection of new training data (as well as the continual update of models) expensive or impossible. In this regard, transfer learning can utilize the limited data between different application domains to model agents' payoff functions in diverse markets; the way is to shape the rewards by learning the increments based on the baselines of classic economic utility models.

When building a system containing machina economicus, the designers should decide which physical components must be embedded and which considerations or features should be integrated into the interaction and learning in a $D^3$ environment. The entities in EI are operating separately and concurrently as autonomous decision-makers who are motivated by their own rationality and not controlled by other entities or a system-wide authority. The information for decision-making is normally scattered across agents, while no one owns a global view. Machina economicus, coined EI in $D^3$ environments, possesses the following challenges arising from the heterogeneity of interactions and the complexity of application requirements:

- **Integration of AI's Selfishness**. For prosumers who aim to learn an efficient and common good policy, an appropriate elicitation and incorporation of AI's economic rationality and strategy is the key, as AIs may behave strategically to advance their benefits and own complex preferences in collective decision-making. The concerns for fairness and trust issues arise when the strategic behaviors are in tension with the decision maker's goals and desires. Moreover, the amount of data generated by EI is increasing very fast with the era of big data.
- **Interdependence and Interaction with $D^3$ Environments**. EI components are no longer isolated as IoT connects various subsystems; instead, EI is driven by multiple agents, each with their own complicated motivations and actions. Machina economicus living in $D^3$ environments gradually loses the luxury of offline thinking while other agents wait for it to make the best decisions. In terms of this, the physical environment becomes nonstationary from the perspective of any individual decision-makers, making the learning or optimization associated with information gathering interdependent.

- **System Modelling**. The straightforward way to model engineering problems is mathematical programming. Unfortunately, most operational problems are highly non-convex and NP-hard, requiring expensive computational efforts to generate global-optimal solutions. Moreover, the application bottlenecks, such as stochastic, time-varying, heterogeneous parameters, scalability, computation tractability, and privacy concerns, pose additional challenges to optimizing, modeling, analyzing, and enhancing large-scale machina economicus.

Whenever machina economicus is deployed with social and economic processes, concerns for system efficiency and social welfare arise when competition, collaboration, or cooperation are in tension with selfish AI's goals and desires. The trade-offs between efficiency and welfare remain the biggest open issues in economics and the related fields. Welfare economics exists as a branch of economics which is concerned with what public policies ought to be, how to maximize individual well-being, and what types of outcomes are preferable (Hu & Chen, 2020). Utility-based notions of welfare can capture the relative benefits that a social good can have on a particular individual.

To better illustrate how machina economicus works, we take the autonomous vehicle (AV), also known as a self-driving car, as an application example. AVs represent the tendency of vehicle-to-Internet in ITS, where AVs and cyber-physical systems should be operated together collaboratively to explore the future electrified and intelligent transportation in greater depth. AV installs physical sensors, actuators, and computers. On top of the advances of self-driving technology, machina economicus could be embodied into AVs to get them involved in the EI. In such scenarios, AVs should not only decide where to steer or when to brake but also interact and collaborate with other vehicles or EIs in a more rational and strategic manner. A future implementation of AVs could be Robo-taxi, which is expected to serve trip demands, determine the optimal route and price for trips, plan the best charging/discharging schedules, and communicate with other vehicles. Robo-taxi is a selfish actuator who may bid for orders, customize trip rates for customers by researching their habits and building profiles, inject power to microgrids as a demand responder, charge other depleted vehicles, and provide computing offloading services, among many other functionalities. Through interacting with ITS, reasoning about other vehicles, and collecting historical/real-time data, Robo-taxi can learn its optimal operational strategy and make online decisions that achieve maximal profits as well as system efficiency during economic interactions with EI or other taxis.

## 4. System Modeling and Methodologies

The management and operations of EI raise various challenges in different integrated subsystems or scenarios, such as data mining, resource allocation, optimization, interaction, and negotiation. The operation issues for machina economicus incorporate three synergistic areas: data analysis and processing, mathematical problem formulation, and intelligent decision-making. In terms of research focuses and challenges, we outline the features and opportunities of the current approaches and advanced technologies in EI. A key to making these technologies efficient in an economic-AI structure is to design proper incentives and policies for selfish AIs in optimization and decision-making, such that EI's sustainability, efficiency and reliability can be improved.

In this section, we will introduce several useful tools that include mechanism design, machine learning and optimization methods, as well as their adaptability to EI management and operation issues. Then we present the proposed integrated solution framework that takes advantage of these AI techniques.

### 4.1 The Advances of AI Techniques
### 4.1.1 Mechanism design

Mechanism design, a field in economics and game theory, takes an objectives-first approach to design economic mechanisms or incentives toward desired outcomes that capture agents' rationality in economic settings. A market-based mechanism allocates goods and services that are difficult to sell traditionally. However, the payoff-relevant types of agents are subject to stochastic changes, and it is also important to keep track of agents' marginal contributions to social welfare over time. Mechanism designers should advance to optimize and automate the overall system and economic processes. The key is to design specified mechanisms with interaction rules, integrated with optimization and machine learning techniques to trade-off the system efficiency and individual well-being.

To this end, market-based mechanisms are able to integrate various machine learning algorithms to accommodate a variety of dynamic settings across periods and agents' changing preferences (Sandholm, 2003). However, critics argue that the preferences and bidding strategies of the selfish homo economicus may vary or change frequently during the bidding process in a decentralized and highly dynamic market environment; thus, it is occasionally not feasible or possible for agents to calculate or adopt such complex bid strategies in real-world scenarios. Considering bidding as a sequential decision, the bidding process can be formulated as a reinforcement learning-to-bid problem (Cai et al., 2017). The goal is to derive the optimal bidding policy through learning. By modelling an auction as a multi-layer neural network, the mechanism design problem can be framed as a constrained learning problem and solved by standard pipelines (Dütting et al., 2019). Allowing the agent feedback on reinforcement learning and using this feedback to define tasks can enable the learning of much more complex behaviors and interactions in various scenarios.

**4.1.2 Machine learning**

Machine learning is widely applied to analyze data and design autonomous systems to adapt to $D^3$ environments. Although deep neural networks present great potential in dealing with large data sets, their ability to learn new concepts quickly, as one of the defining aspects of human intelligence, is quite limited, and meta-learning has been suggested as one strategy to address (Huisman et al., 2021). Moreover, multiagent reinforcement learning (MARL) is a deliberate framework to model machina economicus in a nonstationary and decentralized environment, which induces new modelling approaches characterizing the interconnection of smart structures and necessitating reinforcement learning algorithms for decision-making and optimal management. When there are many agents, they cannot differentiate valuable information that makes cooperative decisions from globally shared information. Besides, each agent's policy changes as training progress, making the environment nonstationary from the perspective of any individual agent. Effective reinforcement learning algorithms were developed that take advantage of multiagent policy-gradient learning based on actor-critic methods (Lowe et al., 2017), the mean-field theory (Yang et al., 2018), or the attention mechanism (Iqbal & Sha, 2019).

For updating a decentralized AI model without user privacy disclosure, federated learning remains the training data distributed over a large number of agents. Each agent computes an update to the current model independently based on its local data and communicates this update to a central controller (Konečn\`y et al., 2016). The key idea is to update the local data and report only a model update to the central controller. To integrate human rationality into federated learning, Toyoda, K. & Zhang (2019) proposed a federated learning-based mechanism design framework to introduce repeated competition such that the selfish agents will follow the protocol with the objective of maximizing their profits. The proposed competitive mechanism in blockchain application indicates that the selected model updates submitted by the agents in the previous round can increase the chance to be voted and more rewards in the next round.

Despite the advantages of various machine learning techniques, the reality is that AI could be strategic in sharing information with others in order to advance its own benefits, either short-term or long-term. The challenges would be how to design machina economicus systems that can learn and generalize from reward signals in unconstrained domains. To this end, the prospect of an economy of AIs has inspired expansions to new mechanism design settings, where AIs are provided with economic incentives and encouraged, as not only a policy-learner but also a policy-maker, to jointly coordinate the operation of EI components and pursue their own benefits that are aligned with the social good. The IoT surroundings motivate us to perform sufficient explorations for designing machina economicus systems and the related interaction or negotiation protocols, which decide when, what information and with whom to communicate, such that agents can learn and generalize from strategic interactions in unconstrained domains with the information asymmetries and dynamics being well tackled.

**4.1.3 Data-driven optimization**

The advances in machine learning and data analysis have intensified the interest in expanding the role of optimization, addressing the problems that are computationally expensive and that are not well mathematically defined (Bengio et al., 2021). Therefore, the multi-source heterogeneous data from EI collected in simulations, physical experiments, production processes, or IoT environments can be better utilized for decision-making and optimization (Jin et al., 2018). The data-driven optimization overcomes the unfitness of the assumptions about the agent's perfect information and precise distribution of uncertainty in centralized mathematical models; instead, decision-makers have historical/real-time data or prior structure knowledge of the probability for optimization.

A close-loop data-driven optimization framework allows feedback from mathematical programming to machine learning, where learning can extract useful and relevant information from data. Mathematical programming can derive optimal solutions from the extracted information. Reversely, machine learning can also guide the searching process by, for instance, learning to branch to devise a model that is both computationally inexpensive and accurate for branching (Gupta et al., 2020) or predicting a representative scenario to obtain near-optimal solutions for two-stage stochastic integer programming (Bengio et al., 2020). In addition, Bengio et al., (2021) used machine learning to build combinatorial optimization algorithms which decompose the combinatorial problem into smaller and simpler learning tasks. Branke et al., (2008) introduced the preference information from a decision maker and interactive methods in the multiple criteria decision-making of multi-objective optimization. Such interactive methods assume that decision-makers have knowledge about the problem but not a deep understanding of optimization theory.

**4.2 Solution Framework**

Machina economicus can learn from each other and the surroundings (M. J. Wooldridge & Jennings, 1995). However, the heterogeneity and selfish nature of machina economicus make the interactions complex, discontinuous, discrete, or nonlinear in a $D^3$ environment. The most critical theoretical and practical issues for the techno-economic analysis and modelling design should consider three perspectives: (1) the use of mathematical formalism to represent and reason about the machina economicus in EI; (2) The architecture modelling is concerned with the design of a strategical and physical system that satisfies the properties specified by machina economicus and EI modules; (3) The interaction protocols or principles are designed and integrated with management and operations schemes in EI. An integrated solution framework that involves AI techniques is depicted in Fig. 1. Recently, several advanced studies have centered on using AI to improve prosumer's benefits, such as automatic bidding (Ren et al., 2017), learning-based mechanism design (Balcan et al., 2005), MARL-based decision-making (Hou et al., 2023), or mechanism design-based AI negotiations (Schulze-Horn et al., 2020).

Developing an integrated solution framework consists of three aspects:

- **Data collection and processing**. The systems operate separately and concurrently to generate massive multi-source heterogeneous data from, for instance, bidirectional charging/refueling stations, smart meters, intelligent vehicles, renewable energies, etc. The data can be evaluated and processed by supervised or unsupervised learning methodologies to extract or predict useful and relevant information.
- **System integration and management**. EI connects various systems to deliver satisfactory energy trades and optimize energy utilization. After collecting and processing the data, the interaction protocols or principles are designed and integrated with the conflicting goals among different entities. The design or management problems can be formulated as computational problems from the extracted information in real-time or offline, using optimization and machine learning to derive optimal solutions for which analytical solutions are unavailable. A hierarchical decision-making framework that integrates optimization, control theory and machine learning should be developed for EI planning, scheduling, and operations.

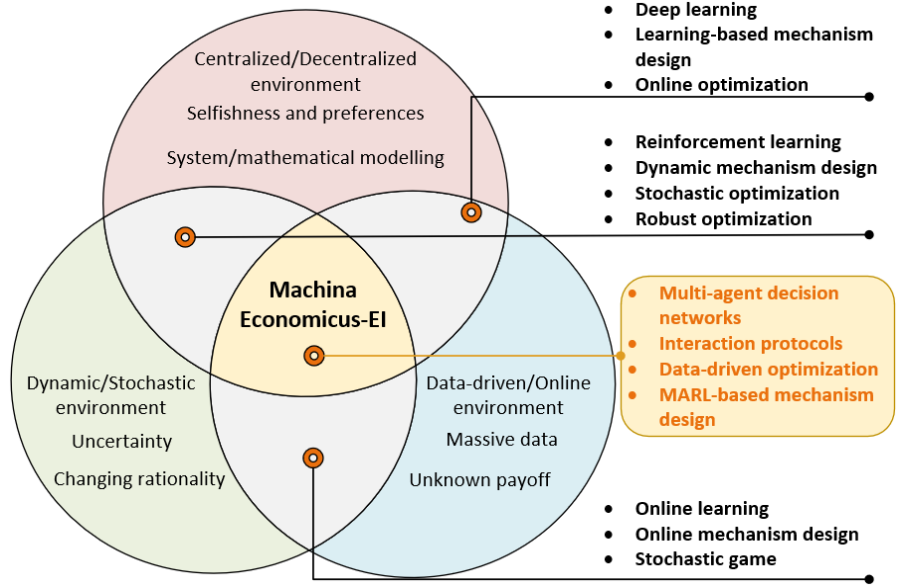

Fig. 1 An integrated solution framework.

- **Digital twins and simulation validation.** The knowledge and model gained from the operations of EI can be applied to arbitrary locations of choice, where more data are integrated, and a larger IoT infrastructure is built. Digital twin or high-fidelity and scalable co-simulation platform could be developed to connect heterogeneous prosumers and promote information and energy exchange in support of Internet-of-Things and 5G communication technologies. By collecting real-time data on EI, a co-simulation platform can also evaluate the behaviors of prosumers in an energy market, and integrate more systems from the real world.

Overall, such an integrated solution for managing and operating large-scale machina economicus systems can be developed such that the objectives related to the physical structures and economic requirements can be fulfilled when facing big data, constraints, and uncertainties in predictions of system status, as well as agent's rationality and selfishness nature. This integrated economic-efficient framework should also incorporate domain knowledge or prediction about EI in the $D^3$ environment and the private information of agents' preferences to enhance operational efficiency and guarantee equilibrium conditions of the whole system.

## 5. Conclusion

In the AI + EI era, new frontiers of network topology, management, and operation schemes in EI are emerging. There remain many open issues and challenges in the machina economicus-centered ecosystem. These questions encourage us to explore and understand how to design and manage EI that captures the economy of AI. The central goal of machina economicus is the design and synthesis of economic, intelligent artifacts in EI. Currently, AI has surpassed humans in some domains but more likely to act to assist human decision-making. We expect the future AI to be capable of thinking and behaving like humans, own more 'economic' features, replace more human side decision-making, and additionally, learn to interact and cooperate with other AIs. Along this trend, our focus has been on designing, implementing, and optimizing the EI, and making it robust against machina economicus's economic rationality and the collective, strategic behaviors, as well as the critical features in $D^3$ environments. This study has covered modern topics and AI techniques and presents novel applications within EI.

The economic-AI perspective we studied gives rise to various challenging issues and encourages applications in EI. For instance, autonomous driving creates a hybrid society comprising vehicles and cyber systems and brings numerous social and economic impacts of V2X coordination to future grid-interactive transportation. With the advances of AI, economic theory has the appealing prospect of widening the applicability of machine learning and optimization techniques to more real-world applications. Achieving system-wide optimality is sometimes inapproachable. Instead, we try to pursue slightly sub-optimal, computational costless beneficial equilibria within decision-making driven by multi-objective optimization, game theory, and machine learning. Additionally, the computational costs of deploying such a framework should be small. By incorporating data from IoT devices, numerical simulations, agent inputs, or physical experiments, machine learning can model and accelerate sequential or online decision-making and address data that is not known analytically for forecasting complex systems under bearable computing resources.

As the research and development of EI are evolutionary, this study presents a broad investigation of the impacts and challenges of adopting economic AI systems, as well as the key research issues and advanced techniques in the realm of EI that represents the economy of AI. Studying and understanding how to build, analyze, model, and manage a machina economicus-centered EI is a strong future research direction, as well as the development of integrated solutions that can deal with the global challenges leading to considerable efficiency, economics, and sustainability, which benefits not just the academia but also the wider society served by EI. Alongside the existing literature, we try to inspire an economic and social outlook to researchers in AI and engineering fields by presenting ideas about what problems exist, how to solve them, and how to treat the new challenges and research issues in the machina economicus-EI infrastructure under investigation.

## Author Biography

**Luyang Hou** is a Research Associate Professor with the School of Computer Science (National Pilot Software Engineering School), Beijing University of Posts and Telecommunications (BUPT), China. He received the Ph.D. degree in information systems engineering from Concordia University, Canada, in 2020. Before joining BUPT, he was a Postdoctoral Research Fellow at the Energy Innovation Laboratory of MéridaLabs, The University of British Columbia from January 2021 to February 2022, and at the Institute for Information Systems Engineering, Concordia University from March 2022 to October 2022. His research interests include operational optimization, mechanism design, and reinforcement learning with applications to Energy Internet, vehicle edge computing, electricity market design, and so on.

**Jun Yan** is an Associate Professor at the Concordia Institute for Information Systems Engineering, Concordia University, Canada. He received the B.Eng. degree in information and communication engineering from Zhejiang University, China, in 2011, and the M.S. and Ph.D. (with Excellence in Doctoral Research) degrees in electrical engineering from University of Rhode Island, USA, in 2013 and 2017, respectively. His research interest includes computational intelligence and cyber-physical security in smart critical infrastructures.

**Yuankai Wu** is a Research Professor at the College of Computer Science, Sichuan University, China. He received the M.S. degree in transportation engineering and the Ph.D. degree in vehicle operation engineering, both from the Beijing Institute of Technology (BIT), Beijing, China, in 2015 and 2019. Prior to joining Sichuan University in March 2022, he was an IVADO postdoc researcher with the Department of Civil Engineering at McGill University. His research interests include intelligent transportation systems, and spatiotemporal data analysis. He was the recipient of the 2022 IEEE Outstanding Paper Award for the IEEE Transactions on Industrial Informatics (TII).

**Chun Wang** is a Professor with the Concordia Institute for Information Systems Engineering, Concordia University, Montreal, QC, Canada. His research interests include the interface between economic models, operations research, and artificial intelligence. His is actively conducting research on multiagent systems, data-driven optimization, and economic model-based resource allocation with applications to healthcare management, smart grid, and smart city.

**Tie Qiu** is a Professor at School of Computer Science and Technology, Tianjin University, China. He was a visiting professor at department of electrical and computer engineering of Iowa State University in U.S. (2014-2015). He serves as an associate editor of IEEE/ACM Transactions on Networking (ToN), IEEE Transactions on Network Science and Engineering (TNSE) and IEEE Transactions on Systems, Man, and Cybernetics: Systems, associate editor of Computers and Electrical Engineering (Elsevier), Human-centric Computing and Information Sciences (Springer). He has authored/co-authored 10 books, over 200 scientific papers in international journals and conference proceedings, such as IEEE/ACM Transactions on Networking, IEEE Transactions on Mobile Computing, IEEE Transactions on Industrial Informatics, IEEE Communications Surveys & Tutorials, INFOCOM, GLOBECOM, etc.